# Toward a Science of Autonomy for Physical Systems

Gregory D. Hager
hager@cs.jhu.edu
Johns Hopkins University

Daniela Rus
rus@csail.mit.edu
Massachusetts Institute of Technology

Vijay Kumar
kumar@seas.upenn.edu
University of Pennsylvania

Henrik Christensen
hic@cc.gatech.edu
Georgia Institute of Technology

Our lives have been immensely improved by decades of automation research – we are more comfortable, more productive and safer than ever before. Just imagine a world where familiar automation technologies have failed.  In that world, thermostats don't work -- you have to monitor your home heating system manually. Cruise control for your car doesn't exist. Every elevator has to have a human operator to hit the right floor, most manufactured products are assembled by hand, and you have to wash your own dishes. Who would willingly adopt that world – the world of the last century -- today? Physical systems – elevators, cars, home appliances, manufacturing equipment -- were more troublesome, more time-consuming, less safe, and far less convenient.

Now, suppose we put ourselves in the place of someone 20 years in the future, a future of *autonomous systems.* A future where transportation is largely autonomous, more efficient, and far safer; a future where dangerous occupations like mining or disaster response are performed by autonomous systems supervised remotely by humans; a future where manufacturing and healthcare are twice as productive per person-hour by having smart monitoring and readily re-tasked autonomous physical agents; a future where the elderly and infirm have 24 hour in-home autonomous support for the basic activities, both physical and social, of daily life.  In a future world where these capabilities are commonplace, why would someone come back to today's world where someone has to put their life at risk to do a menial job, we lose time to mindless activities that have no intrinsic value, or be consumed with worry that a loved one is at risk in their own home?

---

[1] Contact: Ann Drobnis, Director, Computing Community Consortium (202-266-2936, adrobnis@cra.org).
For the most recent version of this essay, as well as related essays, please visit:
cra.org/ccc/resources/ccc-led-white-papers

In what follows, and in a series of associated essays, we expand on these ideas, and frame both the opportunities and challenges posed by autonomous physical systems.

**Automation vs. Autonomy**

Automation has been transforming our world since the industrial revolution. Most of what we experience today in our cars, in our homes, and in our factories is automation; it is not autonomy. How is an autonomous system different than an automated one? The difference between autonomy and automation is subtle, but important. One way to articulate the difference is as follows:

> *Automation* is the implementation of a process to be executed according to a fixed set of rules with little or no human interaction. The automation can be fixed, whereby specific rules are defined for all situations (e.g. an airplane autopilot), or flexible, where different situations (e.g. different manufactured products) are guided by different rules. However, the key idea is that whatever the process is, the rules are defined and fixed in advance to achieve a predetermined outcome under all anticipated inputs. In most cases, the system can effectively be tested against all (or at least a representative set) of inputs to guarantee the desired output.

> *Autonomy* is a property of a system that is able to achieve a given goal independent of external (human) input while conforming to a set of rules or laws that define or constrain its behavior. The key difference is that explicit execution rules are not (and cannot) be defined for every possible goal and every possible situation. For example, an autonomous car will take you to your destination (a goal) or park itself (another goal) while obeying the traffic laws and ensuring the safety of other cars and pedestrians. An autonomous tractor will till a field while avoiding ditches and fences and maintaining safety of the equipment and any human operators. An autonomous bricklaying system will build a wall in many different situations and with many different materials while ensuring the wall conforms to both building plans and building codes.

In short, a key difference is that autonomous systems must be able to act independently and intelligently in dynamic, uncertain, and unanticipated environments.[2] But, no system is omnipotent. Another key element of the science of autonomy will necessarily be that a system must be able to detect when its goals stand in conflict with the laws that govern its behavior, and it must have a way to "fail" gracefully in those situations.

---

[2] Adapted from
http://www.nasa.gov/sites/default/files/atoms/files/2015_nasa_technology_road maps_ta_4_robotics_autonomous_systems.pdf

**The Opportunities of Autonomy**

The opportunities offered by advances in the fundamental science enabling intelligent, autonomous, physical systems are immense, both economically, and socially. Physical capabilities cross nearly every major economic area such as healthcare (which consumes 17% of GDP), manufacturing (7% of GDP), construction (4% of GDP), mining (2% of GDP) or agriculture (1% of GDP)[3]. Enhancing productivity or reducing costs in these areas by enhancing physical intelligence and autonomy would have a dramatic impact on our economy and our wellbeing. Likewise, many important social support systems such as transportation, water and food safety, and disaster aid currently consume many person-hours of time, with no real benefit to the participant or to society at large. In short, developing physically intelligent systems has the opportunity to make society more productive, safer, more livable, and more accessible.

Despite what we see in the popular press, or the latest viral video, achieving this future vision is e*mphatically not within the scope of today's technologies* – it requires substantial advances in both our technical and socio-technical understanding of the science of autonomy. It requires systems that are capable of receiving and carrying out natural language instruction at a relatively high level. It requires systems that can be physically capable in an environment that is unstructured and in situations that were never anticipated or tested. It requires systems that can co-exist with people, and be trusted, safe companions and co-workers.

How can we move forward? Research in autonomy can be framed in two ways -- by looking at how technology developments could create new capabilities, and by understanding how the needs from the relevant applications can frame well-defined concrete problems to be solved. The latter creates a lens through which we can measure the potential impact of advances in autonomy, for example:

(1) **Automated ground transportation:** There are more than 30,000[4] US traffic fatalities each year (10x the deaths on 9/11 or 8x all US casualties in Iraq, *per year*, year after year). Companies working in this area will contribute significantly to the economy, and to the livability of our urban areas. Students trained in this area will accelerate progress, and industry-academic-government cooperation would result from federal funding of research in this area.

(2) **Automated flight systems:** Seamless coordination of satellite data, ground station data, and on-board sensing, would avoid the near-misses and crashes due to human error. Solar-powered permanently flying autonomous aircraft would be helpful in monitoring weather and climate change, and would ensure that sufficient surveillance capabilities exist to locate civilian aircraft locations (e.g., the downed

---

[3] http://www.bea.gov/industry/gdpbyind_data.htm and
http://data.worldbank.org/indicator/SH.XPD.TOTL.ZS
[4] http://www-nrd.nhtsa.dot.gov/Pubs/812160.pdf

Malaysian Air flight from last year). System capacity would be increased to accommodate the expected growth in commercial aviation without adding substantial new infrastructure.

(3) **Disaster response and recovery:** Natural disasters, whether due to weather, disease, or human conflict, require immense mobilization of resources to assess the scope and severity of disaster, to find survivors, to manage the logistics of transport, and to monitor the situation as it evolves. Intelligent distributed cyber-physical resources such as swarms of aircraft, ground vehicles, and fixed equipment could significantly enhance the speed, quality, and scope of disaster response.

(4) **Automation and space exploration:** Robots could be used to harvest material resources from extra-planetary objects or build habitats for people before they arrive. This would in turn enable heavy-launch from low-gravity environments for deep space exploration.

(5) **Automation and agriculture**: Autonomous air vehicles could adapt their own paths to concentrate on problem areas and land-based harvesting equipment could selectively collect ripe items or remove weeds without using pesticides. Water could be used more efficiently, raising yields while reducing environment impact. With such a technology, the US could maintain cheap food, limit exposure of farmers to dangerous fertilizers and pesticides, and reduce incentives for recruiting low-paid illegal immigrants.

(6) **Construction automation**: The time and cost of building a structure could be immensely reduced through enhanced productivity and amplification of construction teams. Site preparation could be automated, increasing quality and decreasing cost. Roofing could be automated, reducing the rate of injury and death due to falls.

(7) **In-home Services:** As the population ages, the opportunity to promote independent living through automation will continue to grow. Walking assistance, assistance with the activities of daily life, activity and health monitoring, and increased social interaction would improve the physical and mental independence and health of the aging population.

(8) **Law Enforcement:** Body-worn cameras will create new opportunities to understand and improve law enforcement methods. Surveillance cameras will allow better tracking of criminal activity and response thereto. Police could be deployed more efficiently and more safely by having improved situational knowledge. Smart transport could slow the flight of criminals and enhance the speed of response.

(9) **Planetary science**: The health and survival of the human species depends on a healthy ocean, healthy forests, and the impact of climate change. Swarms of autonomous robots could help us understand ocean dynamics and the impact of the pollutants on populations of living organisms, the change in forest density, or changes in polar ice coverage. Poor prediction of storm formation and paths lead to poor preparations by communities in the storms path and unnecessary deaths. Swarms of light-weight robots dropped into storms to gather data that would improve storm modeling and prediction.

**The Path Forward**

We are far from having agents that exhibit the breadth of capabilities described above. Why? At a fundamental level, creating physical intelligence is very hard – what we take for granted, for example carefully grasping the arm of an elderly patient to steady them as they rise from a chair, are fantastically difficulty to engineer. Creating resilient systems that can deal with unforeseen situations and untested failure modes is still an emerging science. Imbuing a system with what we consider "common sense" resists even a clear definition, let alone a robust solution. This doesn't even consider the challenges of communication, instruction, or interaction that we expect from co-workers, co-inhabitants, or others we interact with during the course of a normal day.

There is also no question that introducing autonomy will be advantageous to many, but has equal potential to be disruptive. Some types of jobs or activities will be irrevocably altered. Technologies that have the potential to do good will pose new risks. This will lead some to question the wisdom of further automation within our world. As a result, new societal, ethical and legal frameworks will need to emerge or evolve, new types of jobs and roles will be created, and unexpected side-effects and synergies will surprise even the most tech-savvy.

Taken together, these technical and socio-technical challenges frame a number of research questions and challenges, each of which is necessary (but perhaps still not sufficient) to achieve the benefits of physical autonomous systems and to manage the risks:

**Paths to Autonomy**: How are autonomous systems developed? To what extent is autonomy pre-programmed (innate), versus the results of learning, adaptation, and instruction? How do we imbue these systems with capabilities for self-assessment, self-diagnosis, self-organization, and self-repair?

**Engineering of Autonomy:** Is there a science of integration that can inform the engineering of reliable physically autonomous systems? How does the integration of many sub-systems (as is needed for physically intelligent agents) lead to robust intelligence rather than reliability which decreases as function of the failure modes of each new subsystem. How do we ensure safety?

**Sensing and Autonomy:** How do we translate or adapt new ideas in learning to interpret images, videos, or speech signals into methods to adapt grasping from tactile sensing, to detect and adjust the pose of an object to be placed on a shelf, or to react correctly to the movement of a co-worker? Despite tremendous advances in machine perception, reliable, fast, and robust perception remains a major stumbling block for autonomous systems.

**Autonomy and Human Interaction:** How do we create autonomous systems that are perceived as predictable, reliable and trustworthy? How will we interact with autonomous machines that are ubiquitous in society? How will we communicate our intentions to them, and how will they communicate their intentions to us?

**Autonomy and Society:** What are the policy implications of physical autonomy? What are the societal, legal, and ethical issues? What are the economic implications? How do we frame these issues in ways that do not depend on a specific technology or which become rapidly outdated as science and technology evolve?

**Some Closing Thoughts**

It has been nearly 100 years since Karel Čapek penned *Rossum's Universal Robots*[5] and invented a world where technology and the politics of the time came together to intrigue, entertain, and provoke. Today, a century later, our imagination continues to be inspired by the promise of shaping our world through advances in engineered systems, while debating the societal implications of these advances. We contend that, in most cases, the potential human and economic toll of *not* exploring and understanding automation science in a timely and thoughtful manner far outweighs the costs or risks. The associated papers in this series amplify these themes by exploring domains where future advances in the science of autonomy intersects opportunities to advance our collective good.

*For citation use*: Hager G. D., Rus D., Kumar V., & Christensen H. (2015). *Toward a Science of Autonomy for Physical Systems*: A white paper prepared for the Computing Community Consortium committee of the Computing Research Association. http://cra.org/ccc/resources/ccc-led-whitepapers/

*This material is based upon work supported by the National Science Foundation under Grant No. (1136993). Any opinions, findings, and conclusions or recommendations expressed in this material are those of the author(s) and do not necessarily reflect the views of the National Science Foundation.*

---

[5] https://en.wikipedia.org/wiki/R.U.R.